\documentclass[12pt]{article}
\usepackage{graphicx}

\newcommand\pubnumber{NuPhys2018-King}
\newcommand\pubdate{\today}

\def\napoli{Department of Physics and Astronomy\\
University of Southampton, Southampton, SO17 1BJ, U.K.}
\def\support{\footnote{Work supported by the STFC Consolidated Grant ST/L000296/1 and the European Union's Horizon 2020 Research and Innovation programme under Marie Sk\l{}odowska-Curie grant agreements 
Elusives ITN No.\ 674896 and InvisiblesPlus RISE No.\ 690575.}}

\def\Title#1{\begin{center} {\Large #1 } \end{center}}
\def\Author#1{\begin{center}{ \sc #1} \end{center}}
\def\Address#1{\begin{center}{ \it #1} \end{center}}

\newcommand\pubblock{\rightline{\begin{tabular}{l} \pubnumber\\
         \pubdate  \end{tabular}}}
\newenvironment{Abstract}{\begin{quotation}  }{\end{quotation}}
\newenvironment{Presented}{\begin{quotation} \begin{center} 
             PRESENTED AT\end{center}\bigskip 
      \begin{center}\begin{large}}{\end{large}\end{center} \end{quotation}}
\def\Acknowledgements{\bigskip  \bigskip \begin{center} \begin{large}
             \bf ACKNOWLEDGEMENTS \end{large}\end{center}}

\def\beq{\begin{equation}}
\def\eeq#1{\label{#1}\end{equation}}
\def\eeqn{\end{equation}}

\def\beqa{\begin{eqnarray}}
\def\eeqa#1{\label{#1}\end{eqnarray}}
\def\eeqan{\end{eqnarray}}

\let\bar=\overbar

\def\vev#1{\langle #1 \rangle}

\def\Dslash{\not{\hbox{\kern-4pt $D$}}}
\def\dslash{\not{\hbox{\kern-2pt $\del$}}}

\def\msb{{\bar{\ssstyle M \kern -1pt S}}}

\def\th#1#2{\ensuremath{\theta_{#1#2}}}

\def\Dm#1#2{\ensuremath{\Delta m^2_{#1#2}}}

\begin{document}
\begin{titlepage}
\pubblock

\vfill
\Title{Theory Review of Neutrino Models and CP Violation}
\vfill
\Author{ Stephen F King\support}
\Address{\napoli}
\vfill
\begin{Abstract}
Although the measurement of the reactor angle has killed tribimaximal lepton mixing, its descendants survive
with charged lepton corrections, or in less constrained forms such as trimaximal mixing and/or mu-tau symmetry,
each with characteristic predictions. Such patterns may be enforced by a remnant of some non-Abelian discrete family symmetry, possibly together with a generalised CP symmetry, which could originate from continuous gauge symmetry and/or superstring theory in compactified extra dimensions, as a finite subgroup of the modular symmetry.
\end{Abstract}
\vfill
\begin{Presented}
NuPhys2018, Prospects in Neutrino Physics,\\
Cavendish Conference Centre, \\
London, UK, December 19--21, 2018
\end{Presented}
\vfill
\end{titlepage}
\def\thefootnote{\fnsymbol{footnote}}
\setcounter{footnote}{0}
\section{Introduction}
\label{Intro}
Neutrino physics has made remarkable progress since the discovery of neutrino mass and mixing in 1998~\cite{nobel}. 
The reactor angle, unknown before
2012, is now accurately measured by Daya Bay:
$\theta_{13}\approx 8.5^{\circ}\pm 0.2^{\circ}$~\cite{Adey:2018zwh}.
The other lepton mixing angles are determined from global fits
 to be in the three sigma ranges: $\theta_{12}\approx 32-36^{\circ} $ and $\theta_{23}\approx 40-52^{\circ}$,
 with the first hints of the CP-violating (CPV) phase $\delta=125-390^{\circ}$.
The best global fit values with one sigma errors are given in Table~\ref{tab:bfp} \cite{Esteban:2018azc} where 
the meaning of the angles is given in Table~\ref{angles}.

\begin{table}[ht]
\begin{minipage}[b]{0.46\linewidth}
\centering\begin{tabular}{lr}
		\hline
		 &  \multicolumn{1}{c}{NuFIT 4.0} \\
				\hline
		$\th12$ [$^\circ$] & $33.82^{+0.78\phantom{0}}_{-0.76}$ \\
		$\th13$ [$^\circ$] & $8.61^{+0.13\phantom{0}}_{-0.13}$   \\
		$\th23$ [$^\circ$] & $49.6^{+1.0\phantom{00}}_{-1.2}$  \\
		$\delta$ [$^\circ$] & $-145^{+40\phantom{.00}}_{-29}$ \\
		$\Dm21$ [$10^{-5}\rm{eV}^2$] & $7.39^{+0.21\phantom{0}}_{-0.20}$ \\
		$\Dm31$ [$10^{-3}\rm{eV}^2$] & $2.525^{+0.033}_{-0.032}$ \\
		\hline
		\hline
\end{tabular}
\caption{ \footnotesize The nu-fit 4.0 results with one sigma errors without SK atmospheric data for the normal ordered (NO) case,
favoured by current data
\cite{Esteban:2018azc}.
}
\label{tab:bfp}
\end{minipage}\hfill
\begin{minipage}[b]{0.4\linewidth}
\centering
\includegraphics[width=60mm]{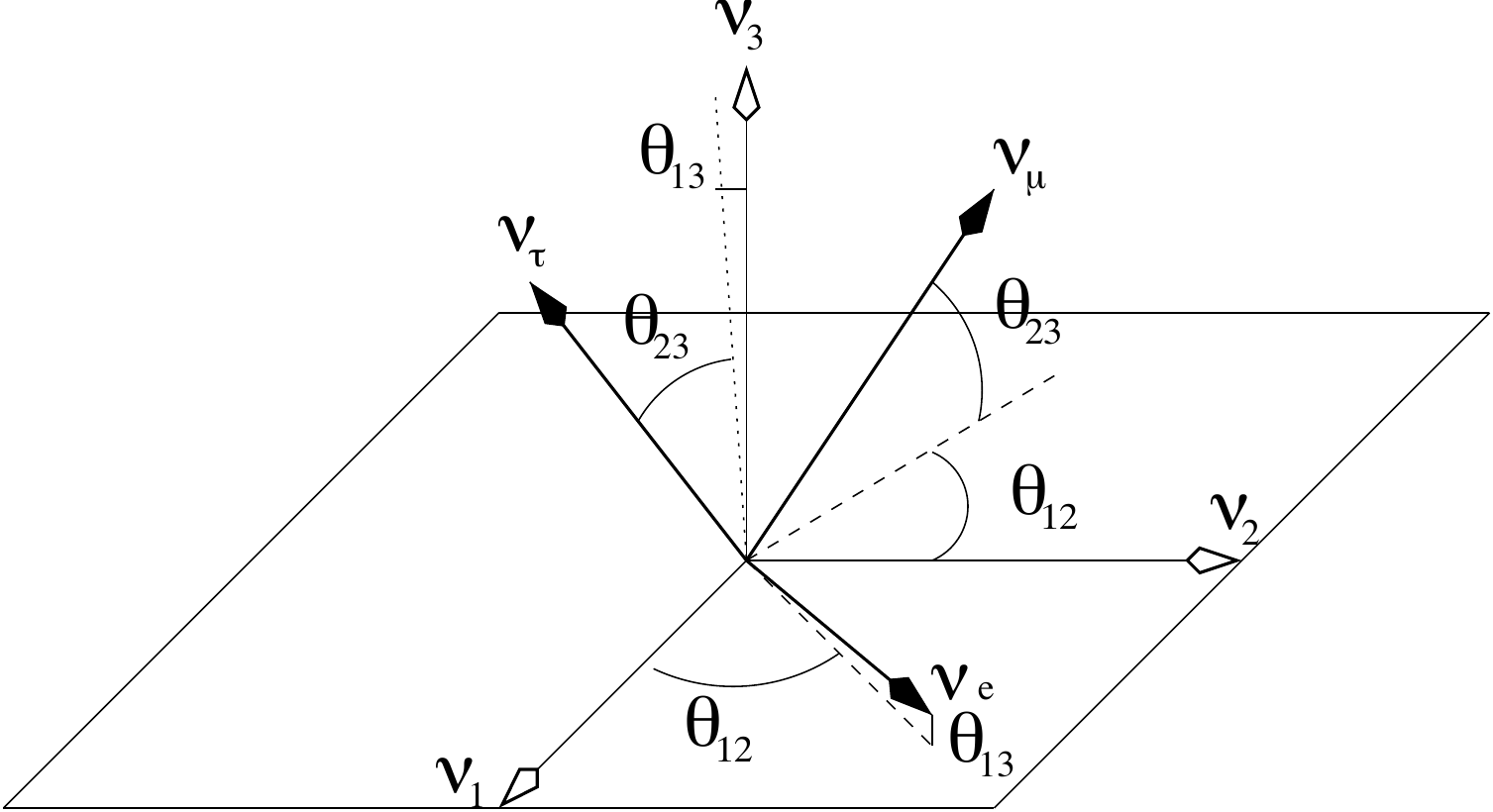}
  \caption{\footnotesize Neutrino mixing angles may be represented as Euler angles relating the states in the charged lepton mass basis
  $(\nu_e, \nu_{\mu}, \nu_{\tau})$ to the mass eigenstate basis states
    $(\nu_1, \nu_2, \nu_3)$.} \label{angles}
\end{minipage}
\end{table}



The measurement of the reactor angle had a major impact on models of
neutrino mass and mixing as reviewed in \cite{King:2013eh,King:2014nza,King:2015aea}
(for earlier reviews see e.g.~\cite{King:2003jb,Altarelli:2010gt,Ishimori:2010au}).
In this talk we give a brief theoretical overview of current neutrino models and CP violation.

\section{Quark vs Lepton Mixing}

The CKM and PMNS mixing matrices are (in the PDG parametrisation):
{\footnotesize 
\begin{eqnarray}
 \label{eq:matrix}
\!\!\!\!\!\!\!\!\!\!\!\!\!\!\!\!\!
\left(\begin{array}{ccc}
    c_{12} c_{13}
    & s_{12} c_{13}
    & s_{13} e^{-i\delta}
    \\
    - s_{12} c_{23} - c_{12} s_{13} s_{23} e^{i\delta}
    & \hphantom{+} c_{12} c_{23} - s_{12} s_{13} s_{23}
    e^{i\delta}
    & c_{13} s_{23} \hspace*{5.5mm}
    \\
    \hphantom{+} s_{12} s_{23} - c_{12} s_{13} c_{23} e^{i\delta}
    & - c_{12} s_{23} - s_{12} s_{13} c_{23} e^{i\delta}
    & c_{13} c_{23} \hspace*{5.5mm}
    \end{array}\right)
   \end{eqnarray}
}
where $s_{13}=\sin \theta_{13}$, etc. with (very) different angles for quarks and leptons.
In the case of Majorana neutrinos, the PMNS matrix also involves the Majorana phase matrix:
$P_M=\textrm{diag}(1,e^{i\frac{\alpha_{21}}{2}},e^{i\frac{\alpha_{31}}{2}})$ which post-multiplies the above matrix.

It is interesting to compare quark mixing, which is small, 
\begin{equation}
s^q_{12}= \lambda , \ \  s^q_{23}\sim \lambda^2, \ \  s^q_{13}\sim \lambda^3
\end{equation}
where the Wolfenstein parameter is $\lambda = 0.226\pm 0.001$,
to lepton mixing, which is large,\footnote{As in section~\ref{Intro} lepton parameters are denoted without a superscript $l$.}
\begin{equation}
s_{13}\sim \lambda /\sqrt{2} , \ \  s_{23}\sim 1/\sqrt{2}, \ \  s_{12}\sim 1/\sqrt{3}.
\end{equation}
The smallest lepton mixing angle 
$\theta_{13}$ (the reactor angle), is of order the largest quark mixing angle 
$\theta^q_{12}=\theta_C=13.0^\circ$ (the Cabibbo angle, where $\sin \theta_C= \lambda$).
There have been attempts to relate quark and lepton mixing angles such as
postulating a reactor angle $\theta_{13}=\theta_C/\sqrt{2}$
\cite{Minakata:2004xt}, and the CP violating lepton phase $\delta \sim -\pi/2$
(c.f. the well measured CP violating quark phase $\delta^q \sim (\pi/2)/ \sqrt{2}$).

\section{Tribimaximal mixing and its descendants}
The tribimaximal (TB) mixing matrix \cite{Harrison:2002er} postulated zero reactor angle $s^2_{13}=\theta_{13}=0$ (hence zero Dirac CP violation), maximal atmospheric angle $s^2_{23}=1/2$ 
($\theta_{23}=45^\circ$) and a solar mixing angle given by $s_{12}=1/\sqrt{3}$ ($\theta_{12}\approx 35.26^\circ$). The mixing matrix is given explicitly by
\begin{equation}\label{TB}
U_{\rm TB} =
\left(
\begin{array}{ccc}
\sqrt{\frac{2}{3}} &  \frac{1}{\sqrt{3}}
&  0 \\ - \frac{1}{\sqrt{6}}  & \frac{1}{\sqrt{3}} &  \frac{1}{\sqrt{2}} \\
\frac{1}{\sqrt{6}} & -\frac{1}{\sqrt{3}} &  \frac{1}{\sqrt{2}}    
\end{array}
\right)P_M,
\end{equation}%
where $P_M$ is the Majorana phase matrix defined above (frequently ignored).
The trimaximal symmetry of the second column,
and bimaximal symmetry of the second and third rows, motivates the use
of non-Abelian discrete symmetries such as A$_4$~\cite{Ma:2001dn}. 
TB mixing was killed~\footnote{Alternative ansatze such as Bimaximal Mixing (BM) and Golden Ratio (GR) Mixing,
not discussed here,  have met the same fate.}
by the measurement of the reactor angle, but it has surviving descendants as follows.

\subsection{Trimaximal lepton mixing and atmospheric sum rules}
\label{atmospheric}
The first surviving descendants of TB mixing are the twins known as 
trimaximal $\rm{TM}_1$ or $\rm{TM}_2$ lepton mixing 
which preserve the first or the second column of Eq.\ref{TB} \cite{Albright:2008rp},
{\small
\begin{equation}\label{TMM}
\!\!\!\!\!\!\!\!
|U_{\rm  TM_1}| =
\left(
\begin{array}{ccc}
\frac{2}{\sqrt{6}} &  - &  - \\ 
\frac{1}{\sqrt{6}} &  - &  - \\
\frac{1}{\sqrt{6}} &  - &  -  
\end{array}
\right),\ \ \ \ 
|U_{\rm TM_2}| =
\left(
\begin{array}{ccc}
- &  \frac{1}{\sqrt{3}} &  - \\ 
- & \frac{1}{\sqrt{3}} &  - \\
- & \frac{1}{\sqrt{3}} &  -  
\end{array}
\right).
\end{equation}%
}
These forms survive since 
the reactor angle becomes a free parameter, while the solar angle may remains close to its
TB prediction (in agreement with data).
The unfilled entries are fixed when the reactor angle
is specified. It is important to emphasise that these forms are more than simple ansatze, since they may be enforced by discrete non-Abelian family symmetry, as discussed in section~\ref{symmetry}.
For example, $\rm{TM}_2$ mixing can be realised by $A_4$ or $S_4$ symmetry~\cite{King:2011zj},
while $\rm{TM}_1$ mixing can be realised by $S_4$ symmetry
\cite{Luhn:2013vna}. A general group theory analysis of semi-direct symmetries 
was given in \cite{Hernandez:2012ra}.

$\rm{TM}_1$ implies three equivalent relations:
\beq
\tan \theta_{12} = \frac{1}{\sqrt{2}}\sqrt{1-3s^2_{13}}\ \ \ \ {\rm or} \ \ \ \ 
\sin \theta_{12}= \frac{1}{\sqrt{3}}\frac{\sqrt{1-3s^2_{13}}}{c_{13}} \ \ \ \ {\rm or} \ \ \ \ 
\cos \theta_{12}= \sqrt{\frac{2}{3}}\frac{1}{c_{13}}
\label{t12p}
\eeqn
leading to a prediction $\theta_{12}\approx 34^{\circ}$,
in excellent agreement with current global fits, assuming $\theta_{13}\approx 8.5^{\circ}$.
By contrast, the corresponding $\rm{TM}_2$ relations imply $\theta_{12}\approx 36^{\circ}$ \cite{Albright:2008rp}, which is on the edge of the three sigma region, and hence disfavoured by current data.
$\rm{TM}_1$ mixing also leads to an exact sum rule relation relation for $\cos \delta$ in terms of the other lepton mixing angles
\cite{Albright:2008rp},
\beq
\cos \delta = - \frac{\cot 2\theta_{23}(1-5s^2_{13})}{2\sqrt{2}s_{13}\sqrt{1-3s^2_{13}}},
\label{TM1sum}
\eeqn
which, for approximately maximal atmospheric mixing, predicts $\cos \delta \approx 0$,
$\delta \approx \pm 90^{\circ}$.~\footnote{
Incidentally the reason why $\cos \delta$ (not $\sin \delta$) is predicted is because such predictions follow from 
$|U_{ij}|$ being predicted, where $U_{ij}=a+be^{i\delta}$, where $a,b$ are real functions of angles in Eq.\ref{eq:matrix}
(hence $|U_{ij}|^2=a^2+b^2+2ab\cos \delta$, which involves $\cos \delta$).}
Such {\em atmospheric mixing sum rules} may be tested
in future experiments~\cite{Ballett:2013wya}. 

For example, the Littlest Seesaw (LS) model~\cite{King:2013iva} leads to $\rm{TM}_1$ mixing, 
for two cases of light Majorana neutrino mass matrix (in the diagonal charged lepton basis):
\begin{equation}
\rm{Case \ I: } \ \ \ \ M^I_\nu= \omega m_a\left(\begin{array}{ccc}
0 & 0 & 0 \\
0 & 1 & 1 \\
0 & 1 & 1
\end{array}
\right)
+
m_s\left(\begin{array}{ccc}
1 & 3 & 1\\
3 & 9 & 3\\
1 & 3 & 1
\end{array}
\right)
\label{I}
\end{equation}
\begin{equation}
\rm{Case\  II: } \ \ \ \ M^{II}_\nu= \omega^2 m_a\left(\begin{array}{ccc}
0 & 0 & 0 \\
0 & 1 & 1 \\
0 & 1 & 1
\end{array}
\right)
+
m_s\left(\begin{array}{ccc}
1 & 1 & 3\\
1 & 1 & 3\\
3 & 3 & 9
\end{array}
\right)
\label{II}
\end{equation}
where $\omega = e^{i 2\pi /3}$.
The LS is very predictive since there are only two free (real) input parameters, 
where $m_a\approx 26$ meV and $m_s\approx 2.6$ meV gives the best fit to neutrino masses 
with $m_1=0$ and PMNS parameters including $\theta_{23}\approx 45^\circ$, $\delta\approx -90^\circ$
(the latter two predictions explained by an approximate mu-tau symmetry as discussed later).

\subsection{Charged lepton mixing corrections and solar sum rules}
\label{solar}

The second way that TB neutrino mixing can survive is due to charged lepton corrections.
Recall that the physical PMNS matrix in Eq.\ref{eq:matrix} is given
by $U_{\rm PMNS}= U^e U^{\nu}$.
Now suppose that 
$U^{\nu}=U^{\nu}_{\rm TB}$, the TB matrix in Eq.\ref{TB}, while
$U^e$ corresponds to small but unknown charged lepton corrections.
This was first discussed in 
\cite{King:2005bj,Masina:2005hf,Antusch:2005kw,Antusch:2007rk}
where the following sum rule involving the lepton mixing parameters, including crucially the CP phase $\delta$,
was first derived (where $35.26^o=\sin ^{-1}\frac{1}{\sqrt{3}}$):
\begin{equation} 
\!\!\!\!\!\!\!\!\!\!
\theta_{12}\approx  35.26^o + \theta_{13}\cos\delta ,
\label{eq:linearSSR} 
\end{equation}

To derive this sum rule, let us consider the case of 
the charged lepton mixing corrections involving only (1,2) mixing,
so that the PMNS matrix is given by \cite{Antusch:2007rk},
{\footnotesize
\begin{equation}
\!
U_{\mathrm{PMNS}} = \left(\begin{array}{ccc}
\!c^e_{12}& s^e_{12}e^{-i\delta^e_{12}}&0\!\\
\!-s^e_{12}e^{i\delta^e_{12}}&c^e_{12} &0\!\\
\!0&0&1\!
\end{array}
\right)
\left( \begin{array}{ccc}
\sqrt{\frac{2}{3}} & \frac{1}{\sqrt{3}} & 0 \\
-\frac{1}{\sqrt{6}}  & \frac{1}{\sqrt{3}} & \frac{1}{\sqrt{2}} \\
\frac{1}{\sqrt{6}}  & -\frac{1}{\sqrt{3}} & \frac{1}{\sqrt{2}}
\end{array}
\right)
= \left(\begin{array}{ccc}
\! \cdots & \ \ 
\! \cdots&
\! \frac{s^e_{12}}{\sqrt{2}}e^{-i\delta^e_{12}} \\
\! \cdots
& \ \
\! \cdots
&
\! \frac{c^e_{12}}{\sqrt{2}}
\!\\
\frac{1}{\sqrt{6}}  & -\frac{1}{\sqrt{3}} & \frac{1}{\sqrt{2}}
\end{array}
\right)
\label{Ucorr}
\end{equation} }
Comparing Eq.~\ref{Ucorr} to the PMNS parametrisation in Eq.\ref{eq:matrix}, we identify
the exact sum rule relations~\cite{Antusch:2007rk}, in terms of the elements 
$|U_{e3}|,|U_{\tau 1}|,|U_{\tau 2}|,|U_{\tau 3}|$
identified above.
The first element $|U_{e3}|=\frac{s^e_{12}}{\sqrt{2}}$ implies a reactor angle $\theta_{13}\approx 9^{\circ}$
if $\theta_e\approx \theta_C$ (see e.g. the models in~\cite{Minakata:2004xt}).
The second and third elements, $|U_{\tau 1}|,|U_{\tau 2}|$ after eliminating $\theta_{23}$, yield
a new relation between the PMNS parameters, $\theta_{12}$, $\theta_{13}$ and $\delta$.
Expanding to first order gives the approximate solar sum rule relations in
Eq.\ref{eq:linearSSR} \cite{King:2005bj}.
The fourth element implies $s_{23}^2<1/2$ which is somewhat disfavoured by global fits.

The above derivation assumes only $\theta^e_{12}$ charged lepton corrections.
However it is possible to derive an accurate sum rule which is valid for both $\theta^e_{12}$ and $\theta^e_{23}$ charged lepton corrections (while keeping $\theta^e_{13}=0$). 
Indeed, using a similar matrix multiplication method to that employed above
leads to 
the exact result  \cite{Ballett:2014dua}:
\begin{equation}
\frac{\left | U_{\tau 1} \right |}{\left | U_{\tau 2} \right |
}=\frac{|s_{12} s_{23} - c_{12} s_{13} c_{23} e^{i\delta}|}
    {|- c_{12} s_{23} - s_{12} s_{13} c_{23} e^{i\delta}|}
= \frac{1}{\sqrt{2}}\;. \label{sol3}
\end{equation}

After some algebra, Eq.\ref{sol3} leads to \cite{Ballett:2014dua},
\footnote{See also~\cite{Marzocca:2013cr} for an earlier derivation based on an analysis of phases.}
\begin{equation}
\cos \delta =\frac
{t_{23}s^2_{12}+s^2_{13}c^2_{12}/t_{23}-\frac{1}{3}(t_{23}+s^2_{13}/t_{23})}
{\sin 2\theta_{12}s_{13}}. \label{sol4}
\end{equation}
To leading order in $\theta_{13}$, Eq.\ref{sol4} returns the sum rule in Eq.\ref{eq:linearSSR},
from which we find 
$\cos \delta \approx 0$ if $\theta_{12}\approx 35^o$,
consistent with $\delta \approx -\pi /2$. 
This can also be understood directly from 
Eq.\ref{sol4} where we see that for $s_{12}^2=1/3$
the leading terms $t_{23}s^2_{12}$ and $\frac{1}{3}t_{23}$ cancel in the numerator, 
giving $\cos \delta = s_{13}/(2\sqrt{2}t_{23})\approx 0.05$ to be compared to  $\cos \delta \approx 0$ in the linear approximation.
In general the error induced by using the linear sum rule instead of the exact one has been shown to be
$\Delta(\cos\delta) < 0.1$  \cite{Ballett:2014dua} for the TB sum rule.
Recently there has been much activity in exploring the phenomenology of various such 
{\em solar mixing sum rules} \cite{Marzocca:2013cr}.
On the other hand, for a GUT example with $\theta^e_{12}\sim \theta_C/3$ and 
$\theta^e_{13}\sim \theta_C$ which violates the {\em solar mixing sum rules} see~\cite{Rahat:2018sgs}.

\section{Family Symmetry Models}
\label{symmetry}
The original motivation for family symmetry was to derive or enforce the TB mixing pattern, or one of the other simple patterns such as BM or GR. These days, the motivation is similar, but applied to one of the surviving descendants of TB mixing.

\subsection{Symmetry of the mass matrices}
The starting point for family symmetry models is to consider the symmetry of the mass matrices.
In a basis where the charged lepton mass matrix $M_e$ is diagonal,
the symmetry is,
\begin{equation}
T^{\dagger}(M_e^{\dagger}M_e)T= M_e^{\dagger}M_e
\end{equation}
where $T={\rm diag}(1, \omega^2 , \omega)$ and $\omega = e^{i2\pi /n}$.
For example for $n=3$ clearly $T$ generates the group $Z^T_3$.
The Klein symmetry $Z^S_2\times Z^U_2$ of the light Majorana neutrino mass matrix is given 
by~\cite{King:2013eh},
\begin{eqnarray}
M^{\nu}= S^TM^{\nu} S, \ \ \ \ M^{\nu}= U^TM^{\nu} U\\
S= U_{\rm PMNS}^*\ {\rm diag}(+1,-1,-1)\ U_{\rm PMNS}^T \label{Sd}\\
U= U_{\rm PMNS}^*\ {\rm diag}(-1,+1,-1)\ U_{\rm PMNS}^T \label{Ud}.
\end{eqnarray}

\subsection{Direct Models}
\begin{figure}[htb]
\centering
\includegraphics[width=0.30\textwidth]{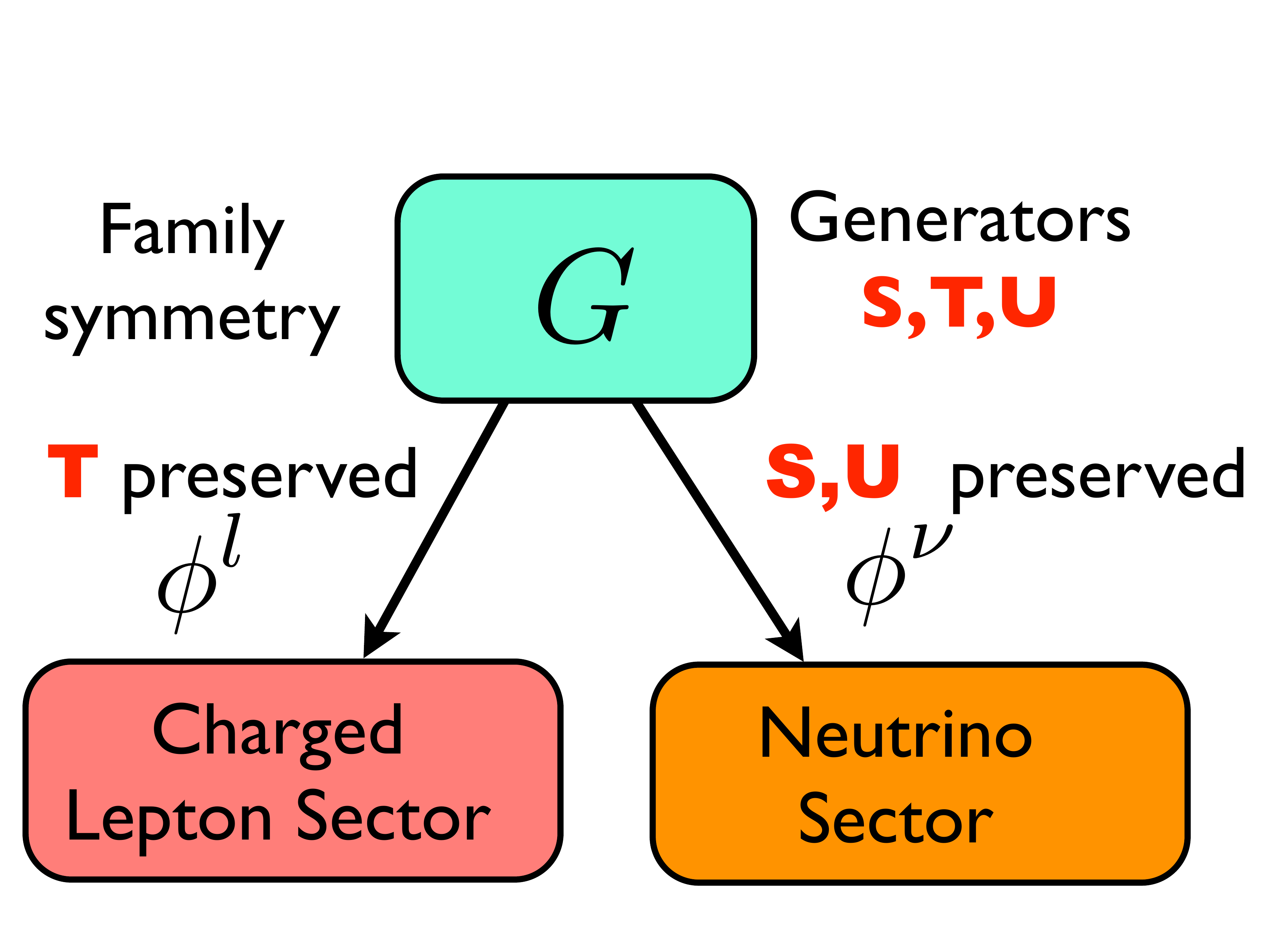}
\includegraphics[width=0.40\textwidth]{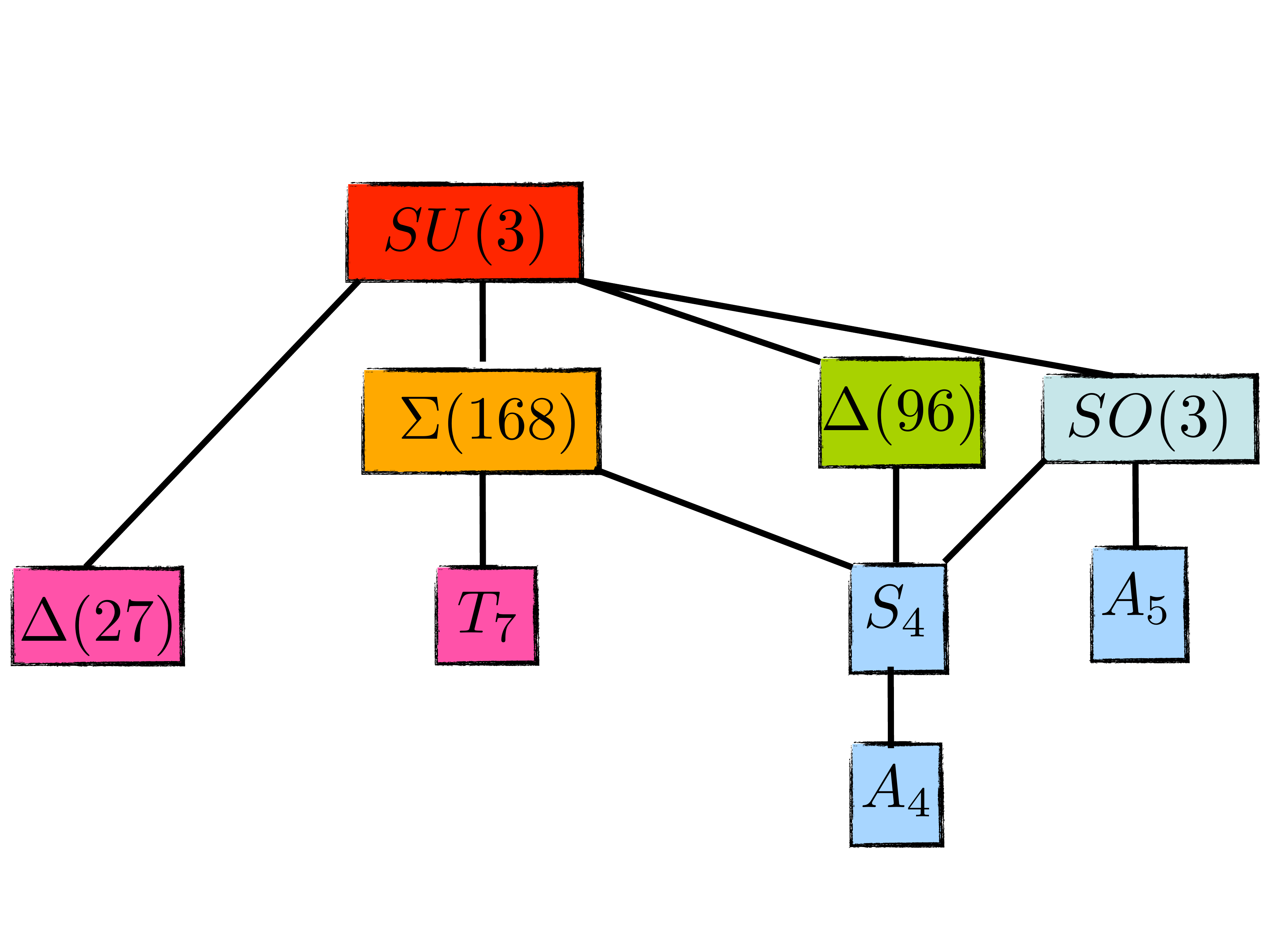}
\vspace*{-4mm}
    \caption{\footnotesize The diagram on the left illustrates the so called direct approach to models of lepton mixing. 
    The diagram on the right shows possible choices of the group $G$.} \label{discrete}
\vspace*{-2mm}
\end{figure}
The idea of ``direct models''~\cite{King:2013eh}, illustrated in Fig.~\ref{discrete} (left panel),
is that the three generators $S,T,U$ introduced above
are embedded into a discrete family symmetry $G$ which is broken by new
Higgs fields called ``flavons'' of two types: $\phi^l$ whose VEVs preserve
$T$ and $\phi^{\nu}$ whose VEVs preserve $S,U$. These flavons are segregated 
such that $\phi^l$ only appears in the charged lepton sector and $\phi^{\nu}$
only appears in the neutrino sector,
thereby enforcing the symmetries of the mass matrices.
Note that the full Klein symmetry $Z^S_2\times Z^U_2$ of the neutrino mass matrix is enforced by symmetry in the direct approach.

There are many choices of the group $G$, with some examples given in Fig.~\ref{discrete} (right panel),
with each choice leading to different lepton mixing being predicted.
For example, consider the group $S_4$ whose irreducible triplet representations are:
~\footnote{There are precise group theory rules for establishing the
irreducible representations of any group, but here we shall only state the results for
$S_4$ in the $T$-diagonal basis, see \cite{Ishimori:2010au} for 
proofs, other 
examples and bases (e.g. dropping the $U$ generator leads to the $A_4$ subgroup).}
\begin{eqnarray}
 \label{eq:matrixU}
\!\!\!\!\!\!\!\!\!\!\!\!\!\!\!\!\!
S=\frac{1}{3} \left(\begin{array}{ccc}
   -1 & 2&2 \\ 2&-1&2 \\2&2&-1   
    \end{array}\right),\ \ \ \ 
       T= \left(\begin{array}{ccc}
    1&0&0\\ 0&\omega^2 &0 \\ 0&0&\omega     \end{array}\right),\ \ \ \ 
   U= \mp \left(\begin{array}{ccc}
     1 & 0 & 0 \\
 0 & 0 & 1 \\
0 & 1 & 0
    \end{array}\right)   
   \end{eqnarray}   
   where $\omega = e^{i2\pi /3}$.
Assuming these $S_4$ matrices,
the $Z^T_3$ symmetry of the charged lepton mass matrix and the Klein symmetry $Z^S_2\times Z^U_2$
of the neutrino mass matrix leads to the prediction of 
TB mixing (indeed one can check that $S$ and $U$ are diagonalised by $U_{TB}$
as in Eqs.\ref{Sd},\ref{Ud}).

\subsection{Semi-direct and tri-direct CP models}
In the ``semi-direct'' approach~\cite{King:2013eh}, in order 
to obtain a non-zero reactor angle, 
one of the generators $T$ or $U$ of the residual symmetry is assumed to be broken.
For example, consider the following 
two interesting possibilities:
\begin{enumerate} 
 \item {The $Z_3^T$ symmetry of the charged lepton
 mass matrix is broken, but the full Klein symmetry $Z_2^S\times Z_2^U$ 
 in the neutrino sector is respected. This corresponds to having 
 charged lepton corrections, with 
 solar sum rules in section \ref{solar}.}
\item{The $Z_2^U$ symmetry of the neutrino mass matrix is broken, but the  
$Z_3^T$ symmetry of the charged lepton
mass matrix is unbroken. In addition either $Z_2^S$ or $Z_2^{SU}$ (with $SU$ being the product of $S$ and $U$) is preserved.
This leads to either $\rm{TM}_1$ mixing (if $Z_2^{SU}$ is preserved~\cite{Luhn:2013vna});
or $\rm{TM}_2$ mixing (if $Z_2^S$ is preserved~\cite{King:2011zj}).
Then we have the atmospheric sum rules as discussed in section \ref{atmospheric}.}
\end{enumerate} 

The ``semi-direct approach'' may be extended to include a generalised CP symmetry $X$
such that $(M^{\nu})^*= X^TM^{\nu} X$, with a separate flavour and CP symmetry in the neutrino 
and charged lepton sectors~\cite{Holthausen:2012dk}
(see also~\cite{deMedeirosVarzielas:2011zw}). Such models typically tend to predict maximal CP violation
$\delta =\pm \pi /2$  (the first example of such generalised CP symmetry is mu-tau reflection symmetry discussed in the following subsection).

In the ``tri-direct'' CP approach~\cite{Ding:2018fyz}, a separate flavour and CP symmetry is assumed for each
right-handed neutrino sector (in the framework of two right-handed neutrino models~\cite{King:1999mb})
in addition to the charged lepton sector.

\subsection{Mu-tau reflection symmetry}
An early class of flavour CP models are based on mu-tau reflection symmetry under which 
$\nu_{\mu}\leftrightarrow \nu_{\tau}^*$ (where the star indicates CP conjugation) leading to the 
prediction of maximal atmospheric mixing $\theta_{23}= \pi/4$ and maximal CP violation
$\delta = \pm \pi /2$~\cite{Harrison:2002et}.
This implies that 
the elements of the second and third rows of the PMNS matrix $U_{PMNS}$
are related by complex conjugation \cite{Harrison:2002et} and 
have equal magnitudes
\cite{Xing:2008fg}; and that 
the elements of the light Majorana neutrino mass matrix $M^{\nu}$ are related~\cite{Grimus:2003yn}. 

For example, the following $\mu\tau$-LS \footnote{The LS refers to 
the fact that these matrices are special cases of the Littlest Seesaw model~\cite{King:2013iva}
in Eqs.\ref{I},\ref{II}
if $m_{a,s}$ are in the special ratio $\frac{m_a}{m_s}=11$
(close to best fit $m_a\approx 26$ meV, $m_s\approx 2.6$ meV).}
light Majorana neutrino mass matrices \cite{King:2019tbt}
\begin{eqnarray}\label{eq:mutau_mass}
M^I_\nu=m_s \left(
\begin{array}{ccc}
 1 & 3 & 1 \\
 3 & 9+11 \omega  & 3+11 \omega  \\
 1 & 3+11 \omega  & 1+11 \omega  \\
\end{array}
\right), \ \ \ \ 
M^{II}_\nu=m_s \left(
\begin{array}{ccc}
 1 & 1 & 3 \\
 1 & 1+11 \omega ^2 & 3+11 \omega ^2 \\
 3 & 3+11 \omega ^2 & 9+11 \omega ^2 \\
\end{array}
\right) \,
\end{eqnarray} 
 where $\omega = e^{i2\pi /3}$, both lead to the PMNS matrix
\begin{eqnarray}
U^{LS}_{\mu\tau}=
\left(
\begin{array}{ccc}
 \frac{2}{\sqrt{6}} & \frac{c_+}{\sqrt{6}} & \frac{c_-}{\sqrt{6}} \\
 \frac{1}{\sqrt{6}} & -\frac{c_+}{\sqrt{6}}-i\frac{c_-}{2} & -\frac{c_-}{\sqrt{6}}+i\frac{c_+}{2} \\
 \frac{1}{\sqrt{6}} & -\frac{c_+}{\sqrt{6}}+i\frac{c_-}{2} & -\frac{c_-}{\sqrt{6}}-i\frac{c_+}{2} \\
\end{array}
\right) 
\end{eqnarray}
which clearly respects $\mu-\tau$ reflection symmetry and is a special case of tri-maximal TM$_1$ mixing \cite{Albright:2008rp} in Eq.\ref{TMM}, 
with a fixed reactor angle $ \frac{c_-}{\sqrt{6}}$, where $c_{\pm} = \sqrt{1\pm \frac{11}{3 \sqrt{17}}}$.
We emphasise that the neutrino mass matrices in Eq.\ref{eq:mutau_mass} have only one free parameter, namely the
neutrino mass scale $m_s$ and so are highly (maximally) predictive! Remarkably, the predicted neutrino masses and PMNS parameters can agree with data after including renormalisation group running effects \cite{King:2019tbt}.

\section{Origin of the non-Abelian discrete symmetry}
While early family symmetry models focussed on continuous non-Abelian gauge theories such as 
$SO(3)$ \cite{King:2005bj} or $SU(3)$ \cite{King:2001uz}, non-Abelian discrete symmetries~\cite{Ishimori:2010au}
are more closely related to TB mixing or its descendants. Here we briefly mention two 
possible origins of such symmetry.

\subsection{Discrete symmetry from continuous symmetry}
It is possible to obtain a non-Abelian discrete symmetry starting from a continuous one
\cite{Koide:2007sr}. For example, we have discussed~\cite{King:2018fke} the breaking of 
supersymmetric $SO(3)$ gauge theory down to $A_4$,
where the $A_4$ may be subsequently broken to smaller residual 
symmetries $Z_3$ and $Z_2$, which may be used to govern the mixing patterns in the charged lepton and neutrino sectors. The basic idea is to use a flavon in the $7$ dimensional representation of $SO(3)$ aligned in a particular direction
to break it to $A_4$, as depicted in Tables~\ref{SO(3)},\ref{7}. Further details are given in~\cite{King:2018fke}.

\begin{table}[ht]
\begin{minipage}[b]{0.4\linewidth}
\centering
\includegraphics[width=60mm]{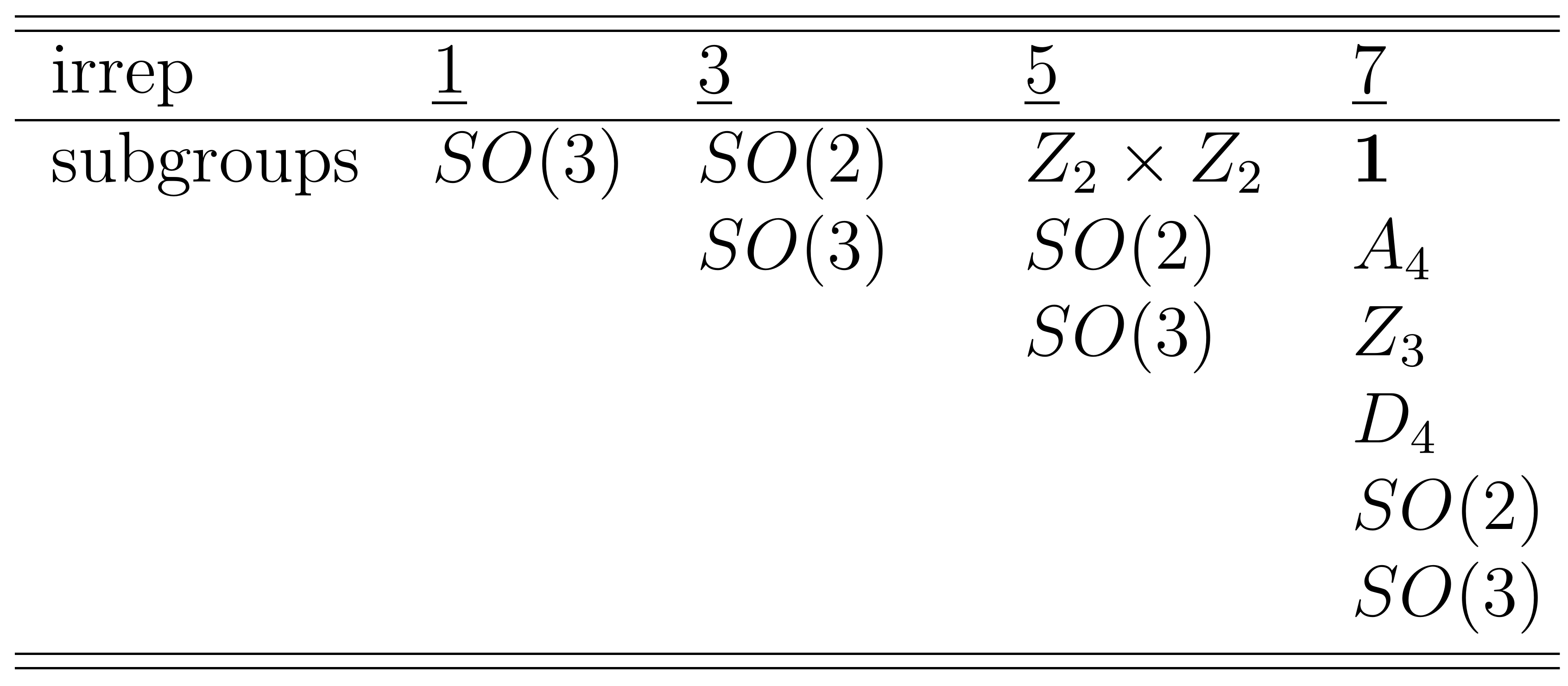}
\caption{ \footnotesize Subgroups of $SO(3)$ preserved when it is broken by flavons in the 
    $1,3,5,7$ dimensional representations of $SO(3)$. 
}
\label{SO(3)}
\end{minipage}\hfill
\begin{minipage}[b]{0.4\linewidth}
\centering
\includegraphics[width=35mm]{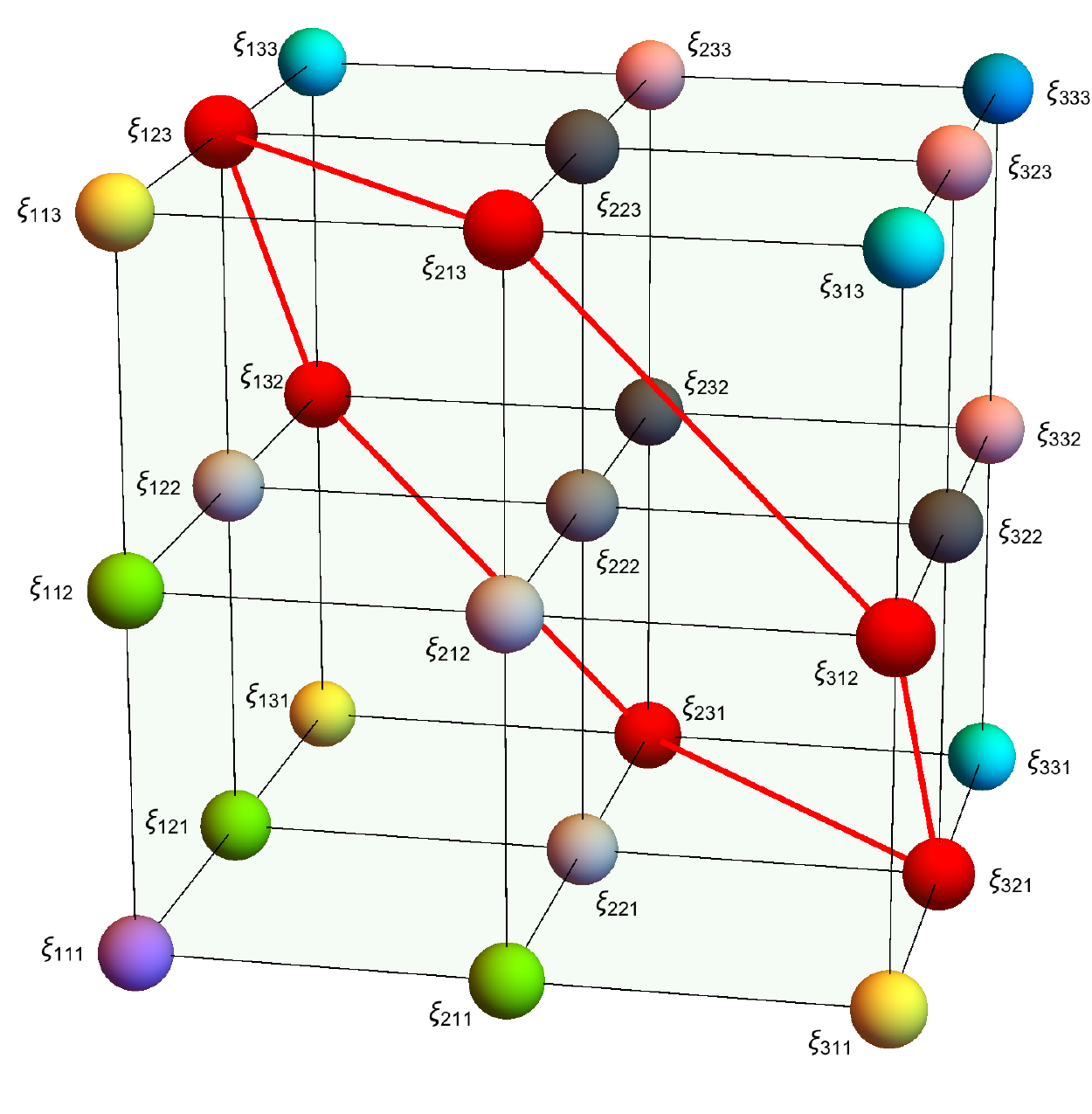}
  \caption{\footnotesize $A_4$ preserving direction of the $7$ dimensional representations of $SO(3)$
        (depicted by the red spheres).}
        \label{7}
\end{minipage}
\end{table}


\subsection{Discrete symmetry from extra dimensions}
\begin{figure}[htb]
\centering
\includegraphics[width=0.32\textwidth]{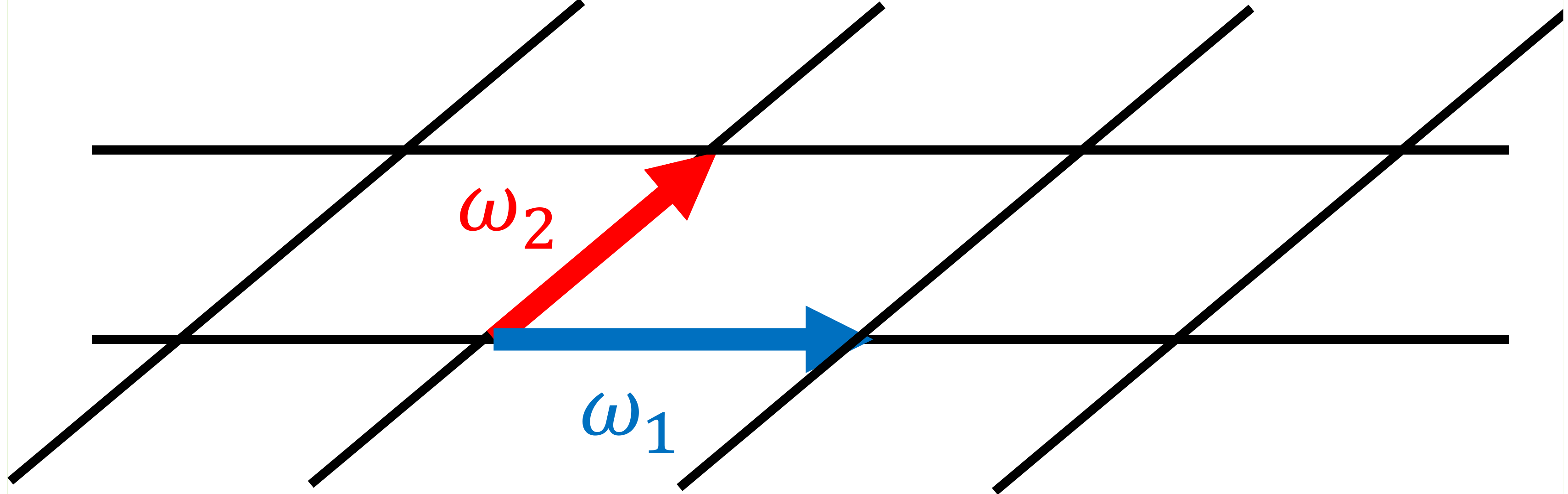}
\includegraphics[width=0.32\textwidth]{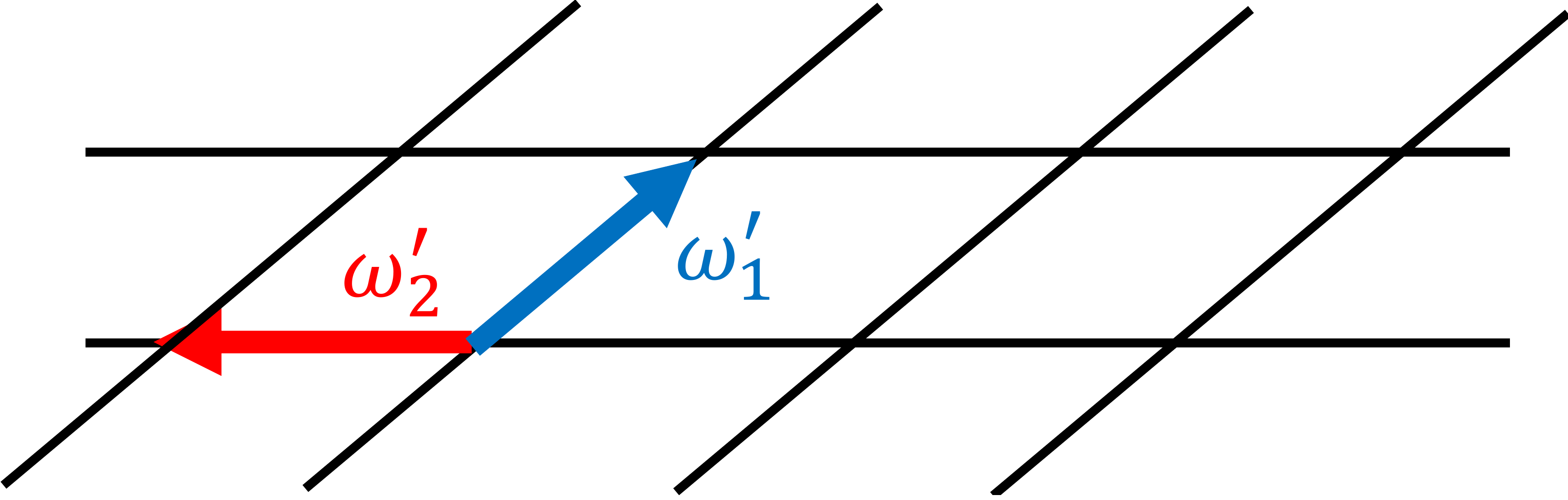}
\includegraphics[width=0.32\textwidth]{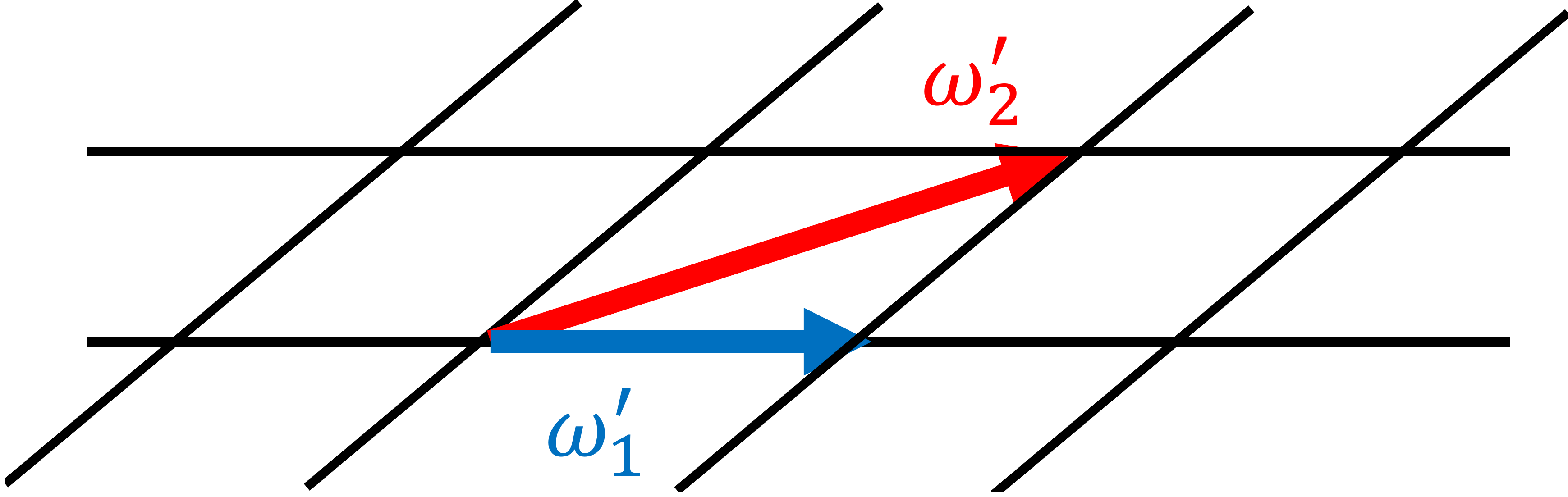}
\vspace*{-4mm}
    \caption{\footnotesize Two extra dimensions compactified on a torus with a twist angle $\theta$ can be represented by a lattice in the complex plane, with basis vectors as shown.} \label{xd}
\vspace*{-2mm}
\end{figure}

Non-Abelian discrete symmetries may arise from superstring theory in compactified extra dimensions,
as a finite subgroup of modular symmetry in the target space~\cite{Altarelli:2005yx,Feruglio:2017spp,Criado:2018thu,deAnda:2018ecu,Novichkov:2018yse}.
Consider a theory with two extra dimensions $x=x_5$ and $y=x_6$ compactified on a torus $T^2$.
If the torus is cut open, its surface is a flat rectangle.
Allowing for a twist angle $\theta$, the torus surface becomes a parallelogram, and an infinite tiling of such parallelograms with identified sides fills the $(x,y)$ (or complex $z=x+iy$) plane to form a lattice structure as shown in Fig.~\ref{xd}.

The lattice is described by two basis vectors $(\omega_1,\omega_2)$ in the complex $z$ plane, as shown in the first panel of~Fig.~\ref{xd}. However the choice of lattice basis vectors
is not unique, and different choices of basis vectors $(\omega'_1,\omega'_2)$ can describe the same lattice. There are two independent transformations on the basis vectors $(\omega_1,\omega_2)$ which leave the lattice invariant as follows. 

The $S$ transformation:
\begin{eqnarray}
\left(
\begin{array}{c}
\omega'_1 \\
\omega'_2\\
 \end{array}
\right)
=
\left(
\begin{array}{cc}
0& 1 \\
-1&  0\\
 \end{array}
\right)
\left(
\begin{array}{c}
\omega_1 \\
\omega_2\\
 \end{array}
\right)
=
\left(
\begin{array}{c}
\omega_2 \\
-\omega_1\\
 \end{array}
\right)
\label{S}
\end{eqnarray}
and the $T$ transformation:
\begin{eqnarray}
\left(
\begin{array}{c}
\omega'_1 \\
\omega'_2\\
 \end{array}
\right)
=
\left(
\begin{array}{cc}
1& 0 \\
1&  1\\
 \end{array}
\right)
\left(
\begin{array}{c}
\omega_1 \\
\omega_2\\
 \end{array}
\right)
=
\left(
\begin{array}{c}
\omega_1 \\
\omega_1+\omega_2\\
 \end{array}
\right).
\label{T}
\end{eqnarray}
The real $2\times 2$ matrices $S$ and $T$ (with $\det S=\det T=1$)
transform the lattice basis vectors as shown in the second and third panels of~Fig.~\ref{xd}. 

Without loss of generality, the lattice can be rescaled as
$(\omega_1,\omega_2)\rightarrow (1,\tau)$, where $\tau \equiv \omega_2/\omega_1$
is a complex modulus field in the upper half of the complex plane 
which describes the compactification~\cite{Altarelli:2005yx}.
The $S,T$ transformations above then apply to the special linear fractional transformations of the modulus field,
$\tau \rightarrow (a\tau +b)/(c\tau + d)$, where $a,b,c,d$ are elements of the matrices $S$ or $T$ above.
Eq.~\ref{S} transforms $\tau \rightarrow -1/\tau$
(associated with compactification radius duality $R\rightarrow 1/R$), while 
Eq.~\ref{T} transforms $\tau \rightarrow \tau +1$, a lattice shift which may be repeated {\it ad infinitum}.
Applying the constraint $T^N=I$, reduces the infinite modular group $\Gamma$
(generated by $S,T$ with $S^2=(ST)^3=I$) into its finite subgroup $\Gamma_N$. 
For example, $\Gamma_3=A_4$,
$\Gamma_4=S_4$, $\Gamma_5=A_5$, are the familiar flavour symmetries~\cite{Altarelli:2005yx}. 

Modular invariance controls orbifold compactifications of the heterotic superstring, hence
the 4d effective Lagrangian must respect modular symmetry.
This implies Yukawa couplings $Y_i(\tau)$ (involving twisted states whose modular weights do not add up to zero)
are modular forms~\cite{Feruglio:2017spp}.
Thus the $Y_i(\tau)$ must form multiplets of $\Gamma_N$, acting 
rather like flavon fields with well defined alignments which depend on $\vev{\tau}$.
In general $\vev{\tau}$ is a free parameter~\cite{Criado:2018thu}, but it may be fixed by the orbifold~\cite{deAnda:2018ecu}
\footnote{Residual symmetries of $\Gamma_3$ for special values of $\vev{\tau}$ have been discussed in 
\cite{Novichkov:2018yse}.}.
For example, a particular orbifold with $\Gamma_3$ and $\vev{\tau}=\omega =  e^{i2\pi /3}$ gives
Yukawa triplet alignments such as 
$Y_i=(-1,2\omega, 2\omega^2)$, respecting 
mu-tau reflection symmetry in the framework of $SU(5)$ Grand Unification~\cite{deAnda:2018ecu}.

\Acknowledgements
I am grateful to the organisers for arranging such a stimulating conference.

\end{document}